\documentclass[%
aip,
 amsmath,amssymb,
 reprint,%
]{revtex4-1}

\usepackage{graphicx}
\usepackage{dcolumn}
\usepackage{bm}

\usepackage[utf8]{inputenc}
\usepackage[T1]{fontenc}
\usepackage{mathptmx}
\usepackage{etoolbox}
\usepackage{hyperref}

\usepackage{xcolor}
\hypersetup{
    colorlinks,
    linkcolor={red!50!black},
    citecolor={blue!50!black},
    urlcolor={blue!80!black}
}

\makeatletter
\def\@email#1#2{%
 \endgroup
 \patchcmd{\titleblock@produce}
  {\frontmatter@RRAPformat}
  {\frontmatter@RRAPformat{\produce@RRAP{*#1\href{mailto:#2}{#2}}}\frontmatter@RRAPformat}
  {}{}
}%
\makeatother
\begin{document}

\preprint{AIP/123-QED}


\title{Prospects for Perovskite/Silicon tandem solar cells to outperform \\ c-Silicon solar cells at elevated temperatures}
\author{Chittiboina Ganga Vinod}
\author{Pradeep R. Nair}
\homepage{Corresponding Author}
 \email{gangavinod@ee.iitb.ac.in, prnair@ee.iitb.ac.in}
 
\affiliation{
Department of Electrical Engineering, Indian Institute of Technology Bombay, Mumbai, India
}

\date{\today}

\begin{abstract}
Successful commercialization of Perovskite/Si tandem solar cells (P/Si TSCs) need a-priori estimation of technological benchmarks to outperform c-Si based technologies under field
conditions. 
To this end, through detailed numerical simulations and analytical modeling, here we identify the limits of ion migration and lifetime degradation till which P/Si TSCs remain competitive. 
 Our results unravel a unique scaling law for the evolution of the efficiency and the temperature coefficient of P/Si TSCs which allows us to anticipate the limiting annual degradation rates. Interestingly, we find that 4T cells are potentially more immune to the ill effects of ion migration as compared to 2T cells. These insights are of broad relevance for material/interface engineering approaches and physics based accelerated tests  which are focused towards long term stability and module reliability. 
\end{abstract}

\maketitle

\section{\label{sec:level1}Introduction}

Commercialization of Perovskite/Si Tandem Solar Cells (P/Si TSCs) appears a promising reality as the efficiencies reported\cite{NREL} ($\approx$ $29.8\%$) are now comparable to the theoretical Shockley-Queisser limits\cite{shockley1961detailed} of single junction silicon solar cells. The economic viability of such tandem solar cells depend on the performance at elevated temperatures in field conditions. Accordingly, the parameters of most importance are the temporal evolution of power conversion efficiency ($PCE$) and temperature coefficient ($T_{PCE}$). As several  phenomena degrade perovskites, it is crucial to identify the phase space of critical parameters where P/Si TSCs retain an efficiency advantage over conventional c-Si solar cells under field conditions.

Efficiency of solar cells at elevated temperatures can be expressed in terms of its temperature coefficient\cite{moot2021temperature} ($T_{PCE}$) as follows 

\begin{equation} \label{eq_1}
PCE(T_H)=PCE(T_R) \times (1+\frac{T_{PCE}}{100}(T_H-T_R))
\end{equation}
            
where  $PCE(T_{H})$ and $PCE(T_{R})$ denote the efficiencies at high temperature ($T_{H}$) and room temperature ($T_{R}$), respectively. Further, let $\alpha=PCE_T(T_R)/PCE_{Si}(T_R)$ denotes the ratio of efficiencies at room temperature, while $\beta=T_{PCE,T}/T_{PCE,Si}$ denotes the ratio of the corresponding temperature coefficients. Note that the subscripts $T$ and $Si$ denote tandem and c-Si, respectively. The corresponding efficiency ratio ($ER$) under field conditions (i.e., at $T_H$) is defined as $ER=PCE_T(T_H)/PCE_{Si}(T_H)$. Simple analysis using eq. \ref{eq_1} indicates that

\begin{equation} \label{eq_2}
ER=\alpha \times \frac{1+\beta \times T_{PCE,Si}(T_H-T_R)/100}{1+T_{PCE,Si}(T_H-T_R)/100} 
\end{equation}

For the economic viability of P/Si TSCs, it is essential that $ER>1$. Interestingly, however, both $\alpha$ and $\beta$ could be time dependent due to the various degradation mechanisms. Hence, the event with $ER=1$ could occur at conditions other than the trivial solution of $\alpha=\beta=1$ which makes long term performance comparisons non-trivial. Under the influence of ion migration and lifetime degradation (due to interface recombination or bulk traps), here we show that 
\begin{equation} \label{eq_21}
\alpha \sim A \beta^{-n} 
\end{equation}

\begin{figure}[ht]
  \centering
    \includegraphics[width=0.45\textwidth]{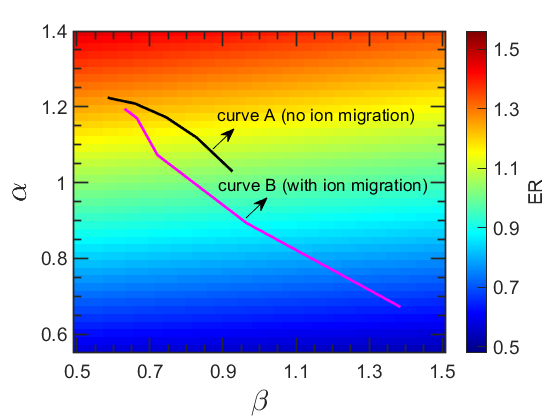}
     \caption{\textit{Phase space of $ER$ as function of $\alpha$ and $\beta$ due to ion migration and interface recombination with $T_H=330K$, with $PCE_{Si}=25\%$ and $T_{PCE,Si}$ = -  $0.3\%/K$. Curve A and B indicate the influence of interface recombination in the absence and presence of ion migration, respectively. Clearly, P/Si TSCs are less competitive in the presence of ion migration.}}
\label{Fig0}
\end{figure}

where $A$ and $n$ are constants influenced by various parameters including the properties of the c-Si bottom cell. Accordingly, Fig. \ref{Fig0} shows that while the P/Si TSCs could remain competitive (i.e., $ER > 1$) in the absence of ion migration (curve A), critical phenomena like interface degradation along with ion migration (curve B) influence the ER in non-trivial manners which could make P/Si TSCs progressively less appealing than c-Si solar cells (i.e., $ER < 1$).

 The rest of the manuscript provides arguments which lead to the results shown in Fig. \ref{Fig0} and the implications of the same. These insights will allow us to identify critical limits of degradation till which P/Si tandem cells remain competitive. Indeed, such quantitative estimates could contribute towards mitigation and characterization strategies (including accelerated tests) from material and interface engineering perspectives.

\section{\label{sec:level2} On the estimation of $\alpha$, $\beta$}
\begin{figure}[ht]
  \centering
    \includegraphics[width=0.45\textwidth]{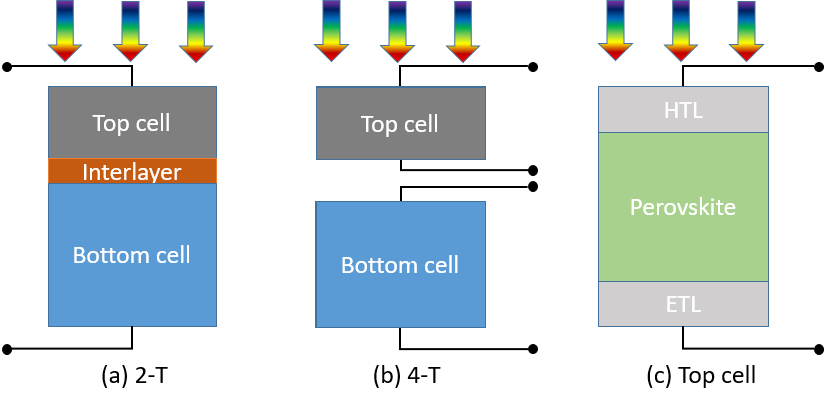}
     \caption{\textit{Schematic of the devices: (a) 2T configuration, (b) 4T configuration, and (c) perovskite top cell.}}
\label{Fig1}
\end{figure}

It is evident from eqs. \ref{eq_2}-\ref{eq_21} that the viability of P/Si TSCs is dictated by the dependence of parameters $\alpha$ and $\beta$ on phenomena like ion migration and interface/lifetime degradation. Modeling the steady state current-voltage characteristics (see Fig.~\ref{Fig1}) with the perovskite top cell under the influence of ion migration is a formidable challenge due to scale of physical mechanisms involved. While the layer thicknesses in the top cell are of the order of hundreds of nm, the ionic screening lengths are of the order of a few nm. On the other hand, the physical dimensions and characteristic lengths of relevant phenomena like diffusion in c-Si bottom cell are of the order of hundreds of $\mu m$. In addition, the optimal architecture for Si bottom cell is still under debate\cite{messmer2021race}. 

Accordingly, here we first obtain the respective steady state J-V characteristics of the top cell and bottom cell with appropriate parameters. As the two cells are connected in series, the J-V characteristics of the tandem configuration is then obtained using the individual J-V characteristics (the validity of this approach was verified using detailed numerical simulations, see Suppl. Mat.). As solar cells are typically employed in maximum power point conditions, the most important parameter is the efficiency under steady conditions. Hence, here we address the steady state efficiency and not the transient or scan rate dependent efficiencies. Hysteresis of tandem solar cells will be addressed in future communications. 

\textbf{Perovskite top cell}: Modeling strategy for perovskite solar cells under the influence of ion migration is well established in prior literature\cite{van2015modeling,kumar2021numerical,courtier2019modelling} as well as our own previous publications 
\cite{nandal2017predictive,saketh2021ion,sivadas2021efficiency}.
Briefly, the electrostatics and transport of charge carriers as well as ions are addressed through the well-known self-consistent scheme of Poisson and continuity equations (see section A and Table S1 of Suppl. Mat. for equations and various parameters involved). Here, we consider negative mobile ions while positive ions are uniformly distributed in the active region. The photo-induced carrier generation rates are estimated through solution of Maxwell's equations using TMM method\cite{burkhard2010accounting}. We consider carrier recombination through trap assisted Shockley-Read-Hall (SRH), Auger, and radiative mechanisms. Further, carrier recombination at the interfaces between perovskite and transport layers are accounted though appropriate recombination velocities\cite{wang2018reducing} ($S$). 

It is well known that band gap of the perovskites  increases with temperature\cite{moot2021temperature,aydin2020interplay}. The same is accounted through an empirical relation, $E_g(T) = E_g(300) + K(T - 300)$ , where K is extracted from the experimental data\cite{moot2021temperature}. In addition, temperature dependence of other parameters like the density of states, capture coefficients, etc. are also considered as per literature\cite{moot2021temperature}. Table II of Suppl. Mat. summarizes the functional dependence of such parameters. Details of the numerical simulation methodology including calibration with experimental results are available in our prior publications 
\cite{saketh2021ion,sivadas2021efficiency,nandal2017predictive,singareddy2021phase}
and in Sec. A-D, Suppl. Mat.

 
  \begin{figure*}[ht]
  \centering
    \includegraphics[width=0.7\textwidth]{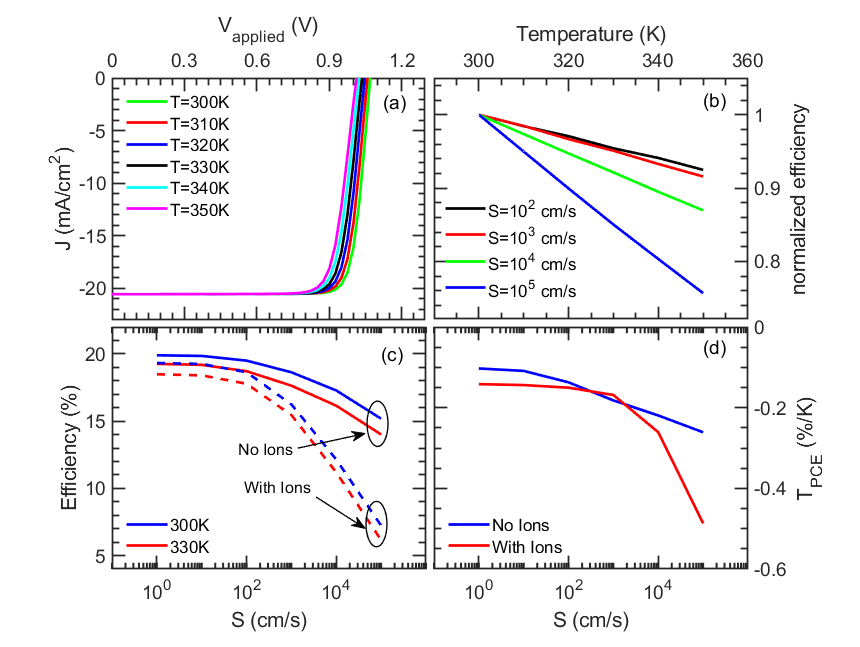}
     \caption{\textit{Top Cell characteristics. (a) J-V for various temperatures under the influence of ion migration (b) Normalised efficiency with respect to Temperature (in the presence of ion migration) (c) Efficiency as a function of interface recombination velocity (S).  (d) Temperature coefficients as a function of $S$. For $S>100$ cm/s, ion migration causes significant degradation of $T_{PCE}$}}
\label{Fig2}
\end{figure*}



 \textbf{Silicon bottom cell}:
 A variety of Silicon cell architectures\cite{messmer2021race} are under consideration for the bottom cell including PERC (Passivated Emitter Rear Cell), TOPCon (Tunnel Oxide Passivated Contact), and other carrier selective designs. However, the optimal configuration among them for tandem cells is yet to be identified\cite{messmer2021race}. Hence, it is important to employ a generic scheme which can anticipate the essential features at elevated temperatures for a variety of bottom cell architectures. In this regard, we describe the bottom cell using the well known ideal diode equation
    \begin{equation}\label{eq:4}
J = -Jsc + J_{0}(e^{(\frac{qV}{\eta KT})}-1) 
    \end{equation}
Here, the parameters $Jsc$, $J_0$ and $\eta$ are appropriately chosen such that the temperature dependent characteristics are well calibrated against typical industry standard c-Si solar cells\cite{green1982silicon,zhao1994reduced}. See Sec. B of Suppl. Mat. for additional details. 


    
    


 \section{\label{sec:results}Influence of ion migration on $\alpha$, $\beta$}
Based on the above described numerical simulation methodology, here we elucidate the functional dependence of $\alpha$ and $\beta$ on critical phenomena like ion migration and interface recombination. To this end, the temperature dependent performance of tandem cell is obtained through the respective characteristics of the top and bottom cells.

\textbf{Perovskite top cell}:
Figure \ref{Fig2} shows the temperature dependent characteristics of the perovskite top cell in the presence of ion migration and interface degradation. Part (a) shows the JV characteristics for various temperatures while part (b) shows that interface degradation critically influences the temperature dependence of normalized efficiency. Importantly, the efficiency degradation is significant for $S > 10^3$ cm/s. 




Part (c) of Fig. \ref{Fig2} illustrates the influence of ions on the efficiency degradation of perovskite top cell. It is evident that the presence of ions exacerbates the influence of interface recombination. Accordingly, the temperature coefficient of efficiency is shown in part (d). 
Interestingly, we find that even for excellent interfaces (i.e., $S<10cm/s$), ion migration degrades the $T_{PCE}$  from  $-0.11\%/K$ to  $ -0.16\%/K$. In the absence of ions, $T_{PCE}$ degrades steadily in the mid range of $S$ (i.e., $1-100$ cm/s). For the same range,  the corresponding variation in $T_{PCE}$ is rather insignificant in the presence of ions. 
For larger $S$ (i.e., $S>100$ cm/s), ion migration is significantly more detrimental and results in rapid degradation in both $PCE$ and $T_{PCE}$. 


\textbf{Silicon bottom cell}:
Temperature dependent J-V characteristics of the Silicon solar cell are shown in Fig. 1, Sec. B of Suppl. Mat.  Current matching conditions with top cell at room temperature is assumed. Accordingly, we use $J_{SC} \sim 20 mA/cm^2$, which is the typical value reported for 2T configurations\cite{aydin2020interplay,de2021efficient}. Similarly, the Voc and FF of our simulations also compare well with results of same publication. In addition, our simulations compare well with the characteristics of typical Si solar cells\cite{green1982silicon,zhao1994reduced} with temperature coefficient of Jsc being $+0.0156 \%/K$ and that of Voc being $-0.22 \%/K$. Accordingly, the $T_{PCE}$ of silicon sub-cell is $-0.29 \%/K$. This indicates that the simple model based on eq. \ref{eq:4} anticipates the temperature dependence of bottom cell. The same approach could be used to explore the effectiveness of a wide variety of c-Si based solar cells as the bottom cell.   

\begin{figure*}[ht]
  \centering
    \includegraphics[width=0.7\textwidth]{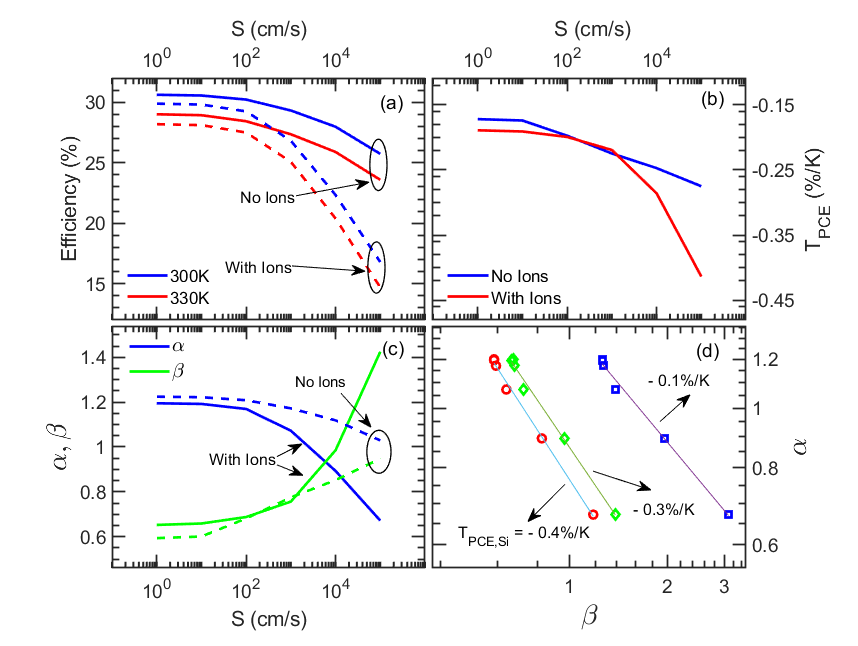}
     \caption{\textit{P/Si 2-T Tandem Cell characteristics. Variation of Efficiency (a) and  temperature coefficient (b) as a function of $S$. Part (c) shows the variation of $\alpha$ and $\beta$ as a function of $S$, with $PCE_{Si}=25\%$} and $T_{PCE, Si}=-0.3\%/K$. Part (d) shows the variation of $\alpha$ as function of  $\beta$ (log-log plot) for various $T_{PCE,Si}$. The trends follow eq. \ref{eq_21}.  }
\label{Fig4}
\end{figure*}

\textbf{2-T tandem cell}:
As mentioned earlier, the characteristics of the tandem cell could be obtained as a combination of the top and bottom cell characteristics  (the validity of this approach was verified using detailed numerical simulations, see Sec. C of Suppl. Mat.).
Accordingly, the influence of interface recombination on  efficiency and $T_{PCE}$ of tandem solar cells, both in the presence and absence of ions, are shown in parts (a) and (b) of Fig. \ref{Fig4}, respectively. While both the top and bottom cell properties influence the $T_{PCE}$ of the tandem cell, the $T_{PCE,T}$ is always better than (or almost equal to) the worst performing sub-cell (see Fig. 4  of Suppl. Mat.). 

Figure \ref{Fig4}b indicates that the $T_{PCE}$ of tandem cells is  of the order of -0.2 to -0.3 $\%/^0K$ for $1<S<10^3$ cm/s. Re-assuringly, the experimentally reported\cite{aydin2020interplay} $PCE$ and $T_{PCE}$  of 2-T tandem cells are of the order of $25\%$ and $-0.26 \%/K$, which compares very well with our simulation results - thus validating the methodology and the parameters involved (see Sec. D of Suppl. Mat. for additional discussion on the comparison with experimental data). Fig. \ref{Fig4}b also indicates that interface degradation beyond $S=100$ cm/s under the presence of ion migration could prove severely critical for P/Si tandem solar cells. 

The results shown in Fig. \ref{Fig4}a,b allow us to gather insights on the degradation dependent variation of $\alpha$ and $\beta$. Part (c) of Fig. \ref{Fig4} indicates that $\alpha$ reduces while $\beta$ increases with an increase in $S$. Interestingly, as the device degrades due to ion migration and interface degradation, we find a unique trend between $\alpha$ and $\beta$ for different $T_{PCE,Si}$ (see Fig. \ref{Fig4}d). Evidently, $\alpha \sim A\beta^{-n}$, with the $n$ being $0.64$, $0.77$ and $0.82$ for $T_{PCE,Si} = -0.1, -0.3$ and $-0.4\%/K$, respectively. This observation forms the basis of eq. \ref{eq_21} which allows us to estimate the competitiveness of P/Si TSC against various c-Si architectures. Similar scaling laws are found to be applicable for various other scenarios like degradation in the carrier lifetime of the perovskite active layer, etc. (details provided in Sec. F of Suppl. Mat.)

\section{Implications}
\textbf{Ion migration and critical $S$}: The results in Fig. \ref{Fig4}c clearly illustrates the ill effects of ion migration. In the absence of ion migration $\alpha > 1$ even for very large $S$. On the other hand, in the presence of significant ion migration, $\alpha$ degrades so much that P/Si TSCs could become noncompetitive.  
It is well known that mobile ions screen the electric field in the perovskite active layer. \cite{calado2016evidence,moia2022dynamics}  As a result, the interface recombination increases with significant ion migration resulting in efficiency loss. Accordingly, it can be analytically shown that the $V_{OC}$ has a $-log(S)$ dependence and hence the variation of $\alpha$ with $S$, as shown in Fig. \ref{Fig4}c. 

Equations \ref{eq_2}-\ref{eq_21} and Fig. \ref{Fig4} allow to estimate the competitiveness of TSCs against - (a) c-Si cells of the same technology as the bottom cell of TSCs, or (b) c-Si cells with a technology other than that of the bottom cells. For part (a), we can rely on eqs. \ref{eq_2}-\ref{eq_21} along with the trends shown in Fig. \ref{Fig4}d. Simple analysis (see Fig. \ref{Fig0}) indicates that P/Si TSCs remain competitive till $\alpha>1$, with a weak dependence on $T_{PCE,Si}$ of the bottom cell. For such cases $S \sim 1000$ cm/s represents the critical limiting interface degradation (as indicated in Fig. \ref{Fig4}c). Sec. G of Suppl. Mat. provides additional theoretical analysis involving eqs. \ref{eq_2}-\ref{eq_21} which leads to the above limits for $S$. 

\textbf{Admissible annual degradation rates}: Direct comparison with other solar cell technologies, for example PERC  solar cells, is possible using eq. \ref{eq_1} itself with the appropriate values of $T_{PCE}$. However, the $T_{PCE}$ of commerical Si solar cells  exhibit a broad range. For example, HIT\cite{Panasonic_corp} (Heterojunction with Intrinsic Thin layer) solar cells report a $T_{PCE}$ of $-0.26\%/K$    while that for PERC\cite{wang2020temperature} solar cells is  $-0.4\%/K$. With such properties for the bottom cell, the results in Fig. \ref{Fig4}c,d  indicate that the $T_{PCE}$ of P/Si TSCs could vary over a range of $-0.25\%/K$ to $-0.5\%/K$. Hence, as per eqs. \ref{eq_1}-\ref{eq_2},  the competitiveness of TSCs is primarily dictated by the parameter $\alpha$. To be competitive over c-Si cells over a period of $N$ years, it can be shown that 

\begin{equation} \label{eq_5}
\gamma_T < \gamma_{Si}+(PCE_T-PCE_{Si})/N
\end{equation}
where $\gamma$ denotes the annual efficiency degradation rate. The subscripts $T$ and $Si$ represents TSC and c-Si solar cells, respectively. A broad range for $\gamma$ is reported for PV modules \cite{jordan2016compendium}. For example, the mean $\gamma$ for c-Si based modules is  $0.6-0.8\%/$ year while the corresponding value for HIT solar cells is $\sim 1\%/$year. As a first order estimate, with $\gamma_{Si} \sim 0.7\%/$year, $N=25$ years, $PCE_{T}=30\%$, and $PCE_{Si}=25\%$, eq. \ref{eq_5} predicts that $\gamma_T < 0.9\%/$ year.  

It is evident that the energy yield of TSCs will be much more than the c-Si over the duration of N years under consideration. Unsurprisingly, improvements in $PCE_{T}$ can offset sub-optimal $\gamma_T$. Interestingly, recent LCOE (Levelized Cost of Electricity) analysis\cite{qian2019impact} indicate similar values for critical $\gamma$ for P/Si TSCs (i.e., $\gamma_T<0.9\%/$year). However, such LCOE analysis seldom correlate degradation rates with any physical mechanisms involved. On the other hand, our analysis clearly anticipate such critical degradation rates along with their correlation to physical degradation phenomena. Further, our analysis also indicates that the current state-of-the art P/Si TSCs (i.e., with $PCE_T \sim 30\%$) have to be nearly as good as c-Si modules (in terms of annual degradation rates) with $S<1000$ cm/s to remain competitive. Evidently, the same results also indicate the need to arrest ion migration for long term stability.

\textbf{Lifetime degradation and  trap density}: The limiting value for $S$ can be, equivalently, interpreted in terms of interface trap density. Bulk trap density of the order of $10^{16} cm^{-3}$ is routinely reported for  perovskites\cite{chen2019facile,ng2020reducing}. With minority carrier lifetimes of the order of $\mu s$, the carrier capture rates are of the order of $10^{-10} cm^{-3}/s$. However, as indicated by recent experiments\cite{al2017new} the interface trap density could be an order of magnitude higher ($\sim 10^{17} cm^{-3}$).  Assuming $nm$ thickness material interfaces, the $S$ is of the order of $1-10$ cm/s for good quality passivated interfaces - an estimate supported by experimental results\cite{grant2017superacid}. Degradation of the interface or lack/loss of passivation could result in $S>1000$ cm/s - another trend supported by literature\cite{yang2015low,wu2017long}. 
 
  In addition to the above, perovskites could have degradation in bulk carrier lifetime ($\tau$). Here, the $\tau$ degradation could introduce different functional dependence on critical parameters like $J_{SC}$, $V_{OC}$, and result in current mismatch between top and bottom cell. Our simulations indicate that eq. \ref{eq_21} holds good for this scenario as well with distinct $n$ and other unique trends (see Sec. F  of Suppl. Mat.).  For example, while both $\alpha$ and $\beta$ vary significantly with $S$, we find that $\beta$ remains rather unvarying with changes in $\tau$. Hence, the time evolution of $\alpha$ and $\beta$ along with $n$ allows one to distinguish between interface vs. bulk degradation of the perovskite active layer.  These insights could be useful to the development of novel characterization strategies.

\textbf{Interface engineering and lifetime prediction}: The non-linear variation of $\alpha$ with $S$ has interesting implications towards interface engineering approaches. 
For example, the trends in Fig. \ref{Fig4}c (and Fig. \ref{Fig2}c) indicate that the $PCE$ remains stable till $S<100$cm/s and rapidly deteriorate afterwards. Hence any interface engineering which results in $S<100$cm/s might seem to be successful towards the challenge of stability improvement - at least during the early days of stability testing. Not surprisingly, such an improvement has been reported by multiple groups \cite{wang2018reducing,wolff2019nonradiative} - although long term stability under field conditions is yet to be thoroughly explored. 

Given the non-linear dependence of $\alpha$ on $S$, it is crucial to accurately characterize the temporal evolution of trap density ($N_T$) near the interface. With $\alpha \propto -log(S)$ for $S>100$cm/s and $S\propto N_T$, it is important to accurately characterize the time evolution of $N_T$. For example, if $N_T$ generation follows a first order process (with various activation energies), the resulting $PCE$ degradation during early years could vary linearly with time. On the other hand, if $N_T$ varies as a power law with time (as seen in interface degradation of field effect devices\cite{alam2007comprehensive}), it  could result in a $-log(t)$ dependence for efficiency degradation (where $t$ is the time). Hence accurate characterization of the time evolution of $N_T$ is crucial to develop predictive models for lifetime and stability.

\textbf{2T vs. 4T}: The methodology described in this manuscript is well suited to compare the effectiveness of 2T configuration vs. 4T configuration. Evidently, 4T configurations are not limited by the current mismatch conditions as compared to 2T schemes. Accordingly, for many scenarios, the theoretical efficiency of 4T cells could be more than that of 2T cells\cite{jiang2016optical,futscher2016efficiency,cheng2021perovskite} - eventhough there are additional system level complexities associated with interconnects and electronics. Our results indicate that for the set of parameters considered in this manuscript, the 4T cells outperform 2T cells. Specifically, the efficiency of 4T cells could be $35\%$ while the corresponding efficiency of 2T cells is $30\%$ (see Sec. H of Suppl. Mat.). We find that ion migration and lifetime degradation influences both configurations in a  similar manner (with comparable $T_{PCE}$). However, the improved baseline efficiency of 4T cells allow them to remain competitive even for larger $S$ (or degradation in $\tau$). The critical $S$, under the influence of ion migration, for 4T cells is $\sim 10^4$ cm/s which is an order of magnitude more than that of 2T cells and hence a significant advantage. In addition, 4T cells might have other advantages like possible replacement of perovskite top layers with additional system level costs.

 In this manuscript, we explicitly considered the influence of inherent or internal degradation mechanisms on tandem cell efficiency while the overall degradation could be influenced by other factors including the bottom cell, encapsulants, metal corrosion, etc., In addition, aspects like interlayer degradation, band level alignments, energy level of traps,  etc. could critically influence the performance of tandem cells (results to be communicated soon including the influence of voltage scan rates). Further, it is evident that phase changes/segregation and associated material degradation could be additional concerns for TSCs.  Indeed, solar spectrum variations (seasonal vs. daily) and ambient climatic conditions (temperature, humidity, etc.) could be other interesting aspects which can influence the overall energy yield and hence the ideal locations favorable for futuristic solar power plants based on P/Si TSCs. Importantly, any deviation from a simple traditional annual linear degradation could have significant implications on methodology to estimate the viability of perovksites and P/Si tandem cells. Clearly more studies are needed in this direction which could result in a synergistic development of predictive models and accelerated tests.

\section{Conclusions}
To summarize, ion migration and lifetime degradation are found to degrade the performance of P/Si TSCs. Through a combination of analytical modeling and detailed numerical simulations, here we identified the phase space of critical parameters which ensure competitiveness of P/Si TSCs. Our results clearly indicate the need to arrest either interface/lifetime degradation or ion migration or both. Further, our methodology allows facile comparison of P/Si TSCs against multiple c-Si technologies and indicates that 4T configuration could be more resilient to the ill-effects due to ion migration. Importantly, our results provide clear technological benchmarks to be met for  P/Si TSCs to outperform c-Si based technologies under field conditions and hence could be of broad relevance to efforts on long term stability and accelerated tests.

\section{Supplementary Materials}
Additional details like simulation methodology, calibration against experimental trends, the influence of $\alpha$ and $\beta$ on bulk lifetime degradation, theoretical arguments which lead to critical $S$, and a comparison between 2T and 4T cells are available in Suppl. Mat. 

\section{Acknowledgements}
This work was funded by Science and Engineering Research Board (SERB, project code CRG/2019/003163), Department of Science and Technology (DST), India. The authors acknowledge IITBNF and NCPRE for computational facilities and Abhimanyu S. for the calibration with TCAD simulations. P.R.N. acknowledges Visvesvaraya Young Faculty Fellowship.

\section*{\textbf{References}}
\bibliography{aipsamp}
\end{document}